\PassOptionsToPackage{table,xcdraw,usenames,dvipsnames}{xcolor}
\documentclass{article} 
\usepackage{iclr2025_conference,times}

\usepackage{amsmath,amsfonts,bm}

\def\eqref#1{equation~\ref{#1}}

\def\1{\bm{1}}

\DeclareMathAlphabet{\mathsfit}{\encodingdefault}{\sfdefault}{m}{sl}
\SetMathAlphabet{\mathsfit}{bold}{\encodingdefault}{\sfdefault}{bx}{n}

\usepackage{hyperref}
\usepackage{url}
\usepackage{amsthm}
\usepackage{ulem}
\usepackage{hyperref}

\usepackage{bm}
\usepackage[most]{tcolorbox}
\usepackage{wrapfig}
\usepackage{multirow}
\usepackage{graphicx,float}
\usepackage{subfigure}
\usepackage{threeparttable}
\usepackage[ruled,vlined]{algorithm2e}
\usepackage{colortbl}
\usepackage{color}
\usepackage{multirow}
\usepackage{tabularx}
\usepackage{float}
\usepackage{graphicx}
\usepackage{booktabs}
\usepackage{arydshln}
\usepackage{enumitem}
\usepackage{wrapfig}
\usepackage{caption}
\usepackage{graphicx}
\usepackage{makecell}
\usepackage{float} 
\usepackage{mathtools}
\usepackage{subfigure}
\usepackage{bbding}
\usepackage{makecell}
\usepackage{tabularx}
\usepackage{amssymb,mathrsfs,amsmath}
\usepackage{pifont}
\usepackage{xcolor}
\usepackage{bbm}

\usepackage{fontawesome5}
\usepackage{siunitx}
\usepackage{listings}

\newcommand{\ourmethod}{{\texttt{MAS$^2$}}\xspace}

\usepackage{twemojis}
\definecolor{mygrey}{gray}{0.4}

\usepackage{cleveref}
\SetKwInOut{Input}{Input}\SetKwInOut{Output}{Output}

\hypersetup{
    colorlinks=true,
    linkcolor=red,
    citecolor=cyan,
    filecolor=magenta,      
    urlcolor=magenta,
    }

\title{MAS$^2$: Self-Generative, Self-Configuring, Self-Rectifying Multi-Agent Systems}

\iclrfinalcopy

\author{Kun Wang$^{\twemoji{deer}\dagger}$
Guibin Zhang$^{\twemoji{lion}\dagger}$,\; 
ManKit Ye$^{\twemoji{bird}\dagger}$,\; 
Xinyu Deng$^{\twemoji{whale}}$,\; 
Dongxia Wang$^{\twemoji{whale}\ddag}$,\\
\textbf{Xiaobin Hu}$^{\twemoji{lion}}$,\;
\textbf{Jinyang Guo}$^{\twemoji{rocket}}$,
\textbf{Yang Liu}$^{\twemoji{deer}}$,
\textbf{Yufei Guo}$^{\twemoji{swan}\ddag}$
	\\
    $^{\twemoji{deer}}$NTU\;
	$^{\twemoji{lion}}$NUS\;
    $^{\twemoji{bird}}$USTC\;
    $^{\twemoji{whale}}$ZJU\;
    $^{\twemoji{rocket}}$BUAA\;
    $^{\twemoji{swan}}$PKU\; $^\dagger$ Equal Contribution\; $^\ddag$ Corresponding\\
\twemoji{email} Main Contact: \texttt{wang.kun@ntu.edu.sg} 
}

\begin{document}

\maketitle

\begin{abstract}
\vspace{-0.7em}
The past two years have witnessed the meteoric rise of Large Language Model (LLM)-powered multi-agent systems (MAS), which harness collective intelligence and exhibit a remarkable trajectory toward self-evolution. This paradigm has rapidly progressed from manually engineered systems that require bespoke configuration of prompts, tools, roles, and communication protocols toward frameworks capable of automated orchestration. Yet, dominant automatic multi-agent systems, whether generated by external modules or a single LLM agent, largely adhere to a rigid ``\textit{generate-once-and-deploy}'' paradigm, rendering the resulting systems brittle and ill-prepared for the dynamism and uncertainty of real-world environments.
To transcend this limitation, we introduce \ourmethod, a paradigm predicated on the principle of recursive self-generation: a multi-agent system that autonomously architects bespoke multi-agent systems for diverse problems. Technically, we devise a ``\textit{generator-implementer-rectifier}'' tri-agent team capable of dynamically composing and adaptively rectifying a target agent system in response to real-time task demands. Collaborative Tree Optimization is proposed to train and specialize these meta-agents. Extensive evaluation across seven benchmarks reveals that \ourmethod achieves performance gains of up to $19.6\%$ over state-of-the-art MAS in complex scenarios such as deep research and code generation. Moreover, \ourmethod exhibits superior cross-backbone generalization, effectively leveraging previously unseen LLMs to yield improvements of up to $15.1\%$. Crucially, these gains are attained without incurring excessive token costs, as \ourmethod consistently resides on the Pareto frontier of cost-performance trade-offs.
The source codes are available at ~\url{https://github.com/yeyeyeah2/MAS2}.

\end{abstract}

\vspace{-1.0em}
\section{Introduction}

\vspace{-0.5em}
Inspired by the collective intelligence observed in human societies~\citep{Book1988_SoM,NeurIPS2023_Agent-SoM}, large language model (LLM)-based multi-agent systems (MAS) have evolved into intricate ecosystems composed of multiple LLMs, tool integrations, memory modules, and communication protocols~\citep{zhuge2024gptswarm,tran2025multiagentcollaborationmechanismssurvey}. In contrast to single-agent paradigms that seek a monolithic, all-powerful model capable of addressing every task in isolation, multi-agent systems harness the virtues of specialization and cooperative interaction among \textit{heterogeneous agents}, with substantial advances across a spectrum of task domains, including scientific discovery~\citep{ghafarollahi2024sciagentsautomatingscientificdiscovery,ghareeb2025robinmultiagentautomatingscientific}, deep research~\citep{zhang2025agentorchestrahierarchicalmultiagentframework,hu2025owloptimizedworkforcelearning}, complex report generation~\citep{yi2025multimodalmultiagentframeworkradiology}, and collective reasoning~\citep{ye2025multiagentsamplingscalinginference}.

The evolution of MAS has unfolded as a clear progression from \textit{manual design} to \textit{full automation}. \textbf{(I)} \textbf{manually configured}: Early frameworks such as AutoGen~\citep{autogen}, MetaGPT~\citep{hong2023metagpt}, and AgentVerse~\citep{chen2023agentverse} relied entirely on hand-crafted specifications, including prompts, agent roles, tool integrations, and communication topologies. \textbf{(II)} \textbf{partially automated}: Subsequent efforts introduced partial automation: for example, GPTSwarm~\citep{zhuge2024gptswarm} and G-Designer~\citep{zhang2024g-designer} automated the design of inter-agent communication topologies, while MasRouter~\citep{yue2025masrouter} and LLM-Selector~\citep{chen2025optimizingmodelselectioncompound} focused on automating LLM routing. \textbf{(III)} \textbf{fully automatic}: Recently, the community has converged on fully automated MAS whose configurations are synthesized end-to-end and adapted online to domain and instance alike, as exemplified by ADAS~\citep{hu2025adas}, MaAS~\citep{zhang2025maas}, and AFlow~\citep{zhang2024aflow}. This shift has inaugurated a phase of rapid, genuinely transformative advance.

Nevertheless, not all autonomously constructed MAS are alike. Early approaches often relied on \textbf{external modules} such as Bayesian optimization~\citep{li2023metaagents}, Monte Carlo tree search (MCTS)~\citep{zhang2024aflow,liang2025imctsenhancingagenticautoml}, or graph neural networks~\citep{zhang2024g-designer}. While effective to some extent, these methods were confined to a predefined search space of atomic operators (\textit{e.g.}, Chain-of-Thought (CoT), Reflexion, Debate), limiting their capacity for architectural innovation. More recently, researchers have shifted toward enabling \textbf{agents themselves} to construct MAS, employing techniques such as SFT~\citep{ye2025masgpttrainingllmsbuild}, DPO~\citep{wang2025scoreflowmasteringllmagent}, and GRPO~\citep{nie2025weakforstrongtrainingweakmetaagent}. These advances allow task-level adaptivity, whereby distinct systems can be customized for each problem instance. Yet most of these methods still adhere to a ``\textit{generate-once-and-deploy}'' paradigm, where a system is instantiated once per task and executed unchanged~\citep{zhang2025maas}, regardless of success or failure. This paradigm proves fundamentally inadequate, as real-world interactions between multi-agent systems and their environments are inherently dynamic and error-prone (\textit{e.g.}, network failures, tool crashes, file loss)~\citep{cemri2025multiagentllmsystemsfail}. Consequently, ``\textit{generate-once-and-deploy}'' strategies are highly susceptible to collapse from a single unforeseen disruption and lack the capacity to adapt beyond their initial instantiation.

Having surveyed the evolution and inherent limitations of both external module-based and agent-driven automatic agent systems, we propose a third new paradigm that inherently enables self-generation and self-adaptation of MAS. Specifically, we envision a framework in which \textbf{a multi-agent system autonomously constructs another multi-agent system}, which we refer to as \ourmethod. Unlike previous approaches, \ourmethod orchestrates a specialized, LLM-facilitated meta-agent team tasked with \textit{generating}, \textit{configuring}, and \textit{rectifying} systems in a task-adaptive and progress-aware manner. By internalizing distinct construction responsibilities across dedicated meta-agents, \ourmethod transcends the creative constraints of external modules and overcomes the rigidity of traditional ``generate-once-and-deploy'' strategies.

In this paper, we introduce a meta multi-agent system (meta MAS) designed to instantiate our proposed \ourmethod paradigm. This system is built upon a tri-agent architecture comprising a \ding{168} \textbf{\textit{generator}}, an \ding{161} \textbf{\textit{implementor}}, and a \ding{171} \textbf{\textit{rectifier}}, each undergoing specialized training to internalize its distinct meta-generative function. Specifically, for any given query, the \textit{generator} agent architects a high-level, multi-agent workflow template, which outlines the sequence of agentic operations. Subsequently, the \textit{implementor} agent instantiates this template by populating each procedural step with a concrete LLM backbone, rendering the workflow fully executable. During runtime, the \textit{rectifier} agent actively monitors the execution state and environmental feedback, issuing timely corrections to the system for adaptiveness to dynamic conditions.

\begin{figure*}[!t]
\centering
\includegraphics[width=\linewidth]{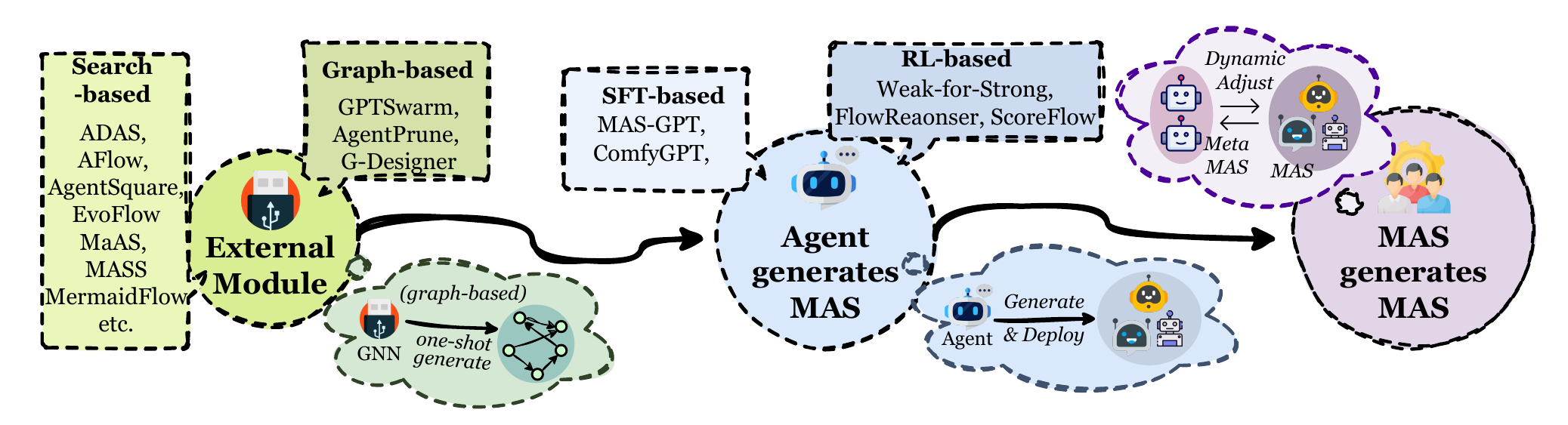}
\vspace{-1.7em}
\caption{The paradigm shift of automatic multi-agent system design: from external module-based MAS generation (\textit{e.g.}, by GNNs, evolutionary algorithms, and search algorithms) and agent-based generation (\textit{i.e.}, FlowReasoner~\citep{gao2025flowreasonerreinforcingquerylevelmetaagents} and MAS-GPT~\citep{ye2025masgpttrainingllmsbuild}) to \ourmethod, where a MAS recursively generates another MAS.}
\label{fig:intro}
\vspace{-2.4em}
\end{figure*}

To facilitate training and data efficiency, we devise an offline reinforcement learning (RL) strategy for trajectory collection and optimization, termed \textbf{Collaborative Tree Optimization (CTO)}. Within the CTO framework, the three agents collaboratively expand a decision tree representing diverse MAS configurations and execution pathways. Upon reaching a terminal state, a final environmental feedback signal is obtained. We then employ a \textit{path credit propagation} mechanism to attribute the final outcome to the upstream decisions made by each agent. This process yields role-specific preference data, which is then leveraged through relative reward-based preference alignment algorithms to effectively specialize the distinct meta-generative function of each agent. {{In this way, \ourmethod materializes a fully autonomous and self-adaptive multi-agent system pipeline, affording robustness and precision commensurate with the demands of intricate, long-horizon task.}}

In brief, our contribution can be summarized as follows:\vspace{-0.5em}
\begin{itemize}[leftmargin=2em,itemsep=-0.1em]
\item[\ding{182}] \textbf{Paradigm Formulation}: We introduce the \ourmethod paradigm, wherein a multi-agent system itself is employed to generate diverse multi-agent systems, thereby achieving greater creativity and task adaptiveness than previous external module or agent-based generation pipelines.

\item[\ding{183}] \textbf{Methodology Proposal}: We instantiate \ourmethod through a ``generator–implementor–rectifier'' tri-agent framework, and design a collaborative tree optimization (CTO) procedure to enable efficient specialization and training of meta-agents.  

\item[\ding{184}] \textbf{Empirical Evaluation}: Comprehensive experiments across six benchmarks demonstrate that \ourmethod achieves \textbf{(I) superior competence}, surpassing state-of-the-art multi-agent systems such as MaAS and ScoreFlow by up to $9.3\%$; \textbf{(II) Pareto-optimal cost–performance}, achieving the highest performance along the cost–performance frontier; and \textbf{(III) cross-backbone generalization}, effectively leveraging unseen LLMs to deliver improvements of up to $15.1\%$.
\end{itemize}

\vspace{-0.8em}
\section{Related Work}
\vspace{-0.6em}
\paragraph{Automating Agent Systems}  concerns multi-agent systems whose configurations (including agent prompting, communication protocols, tool integration, and LLM backbone selection) are automatically determined without substantial human intervention. Early efforts primarily relied on external control modules: for instance, GPTSwarm~\citep{zhuge2024gptswarm} and AgentPrune~\citep{zhang2024cut} employed parameterized adjacency matrices; G-Designer~\citep{zhang2024g-designer} and GraphRouter~\citep{feng2024graphroutergraphbasedrouterllm} utilized graph neural networks; while MasRouter~\citep{yue2025masrouter} adopted a variational autoencoder. A broader class of systems, such as ADAS~\citep{hu2024adas}, AgentSquare~\citep{shang2024agentsquare}, AFlow~\citep{zhang2024aflow}, MaAS~\citep{zhang2025maas}, MASS~\citep{zhou2025multiagentdesignoptimizingagents}, and MermaidFlow~\citep{zheng2025mermaidflowredefiningagenticworkflow}, similarly follow this paradigm.  
However, the generative capacity of external modules proved inherently constrained, giving rise to model-driven approaches for constructing MAS. MAS-GPT~\citep{ye2025masgpttrainingllmsbuild} employs large-scale query–workflow pairs for supervised fine-tuning of a system-generating agent, while ScoreFlow~\citep{wang2025scoreflowmasteringllmagent} adopts a similar strategy under DPO. FlowReasoner~\citep{gao2025flowreasonerreinforcingquerylevelmetaagents} and Weak-for-Strong~\citep{nie2025weakforstrongtrainingweakmetaagent}, leverage GRPO-style online RL~\citep{guo2025deepseek} to train meta-agents for MAS generation.  
In contrast, our proposed \ourmethod transcends the limitations of the ``generate-once-and-deploy'' paradigm by enabling self-adjusting and resilient construction of MAS.

\vspace{-0.8em}
\paragraph{Meta LLM Agents.} Meta agents typically denote LLM-based coordinating entities that provide meta-level guidance or generation for agent teams~\citep{xiong2025mpoboostingllmagents}. Most existing meta agents are instantiated directly from off-the-shelf, powerful LLMs and are pervasive across domains. For instance, in deep research systems, Camel's OWL~\citep{hu2025owloptimizedworkforcelearning}, Skywork's AgentOrchestra~\citep{zhang2025agentorchestrahierarchicalmultiagentframework}, ByteDance's AIME~\citep{shi2025aimefullyautonomousmultiagentframework}, Tencent's Cognitive Kernel-Pro~\citep{fang2025cognitivekernelproframeworkdeep}, and several other frameworks~\citep{bahdanau2024tapeagentsholisticframeworkagent} all rely on a meta agent to allocate and regulate tasks among sub-agents. Analogously, in software engineering systems, manager or ``CEO'' agents are widely adopted~\citep{hong2023metagpt,software-dev,hu2024evomac}. More recently, a smaller body of work has explored explicitly training a meta agent as a leader~\citep{estornell2025howtotrainaleader,gao2025flowreasonerreinforcingquerylevelmetaagents}. Distinct from these efforts, \ourmethod elevates the concept further by transforming the meta agent from a single controlling entity into a \emph{meta-MAS}.

\vspace{-0.8em}
\paragraph{RL for MAS.} RL has been witnessed to propel the agentization of LLMs at an unprecedented pace~\citep{zhang2025landscapeagenticreinforcementlearning}, substantially advancing their capabilities in reasoning, tool utilization, memory management, and beyond. Within the specific context of multi-agent systems (MAS), existing research on RL for MAS can be broadly categorized into three streams: \textit{(i) training external modules}, which leave the internal parameters of agents untouched while exclusively training external control modules via RL~\citep{zhang2025maas,wang2025gsafeguardtopologyguidedsecuritylens,wang2025agentdropoutdynamicagentelimination}; \textit{(ii) training partial agents}, which selectively train a subset of agents while keeping the remainder fixed, as exemplified by MLPO~\citep{estornell2025howtotrainaleader}; and \textit{(iii) comprehensively training all agents} that jointly update and evolve all constituent agents, such as Sirius~\citep{zhao2025sirius}, MALT~\citep{motwani2024malt}, MaPoRL~\citep{park2025maporl}, MARFT~\citep{liao2025marft}, among others~\citep{wan2025remalearningmetathinkllms,wang2025coevolvingllmcoderunit,xia2025mmedagentrloptimizingmultiagentcollaboration,liu2025llmcollaborationmultiagentreinforcement}.

\begin{figure*}[!t]
\centering
\includegraphics[width=\linewidth]{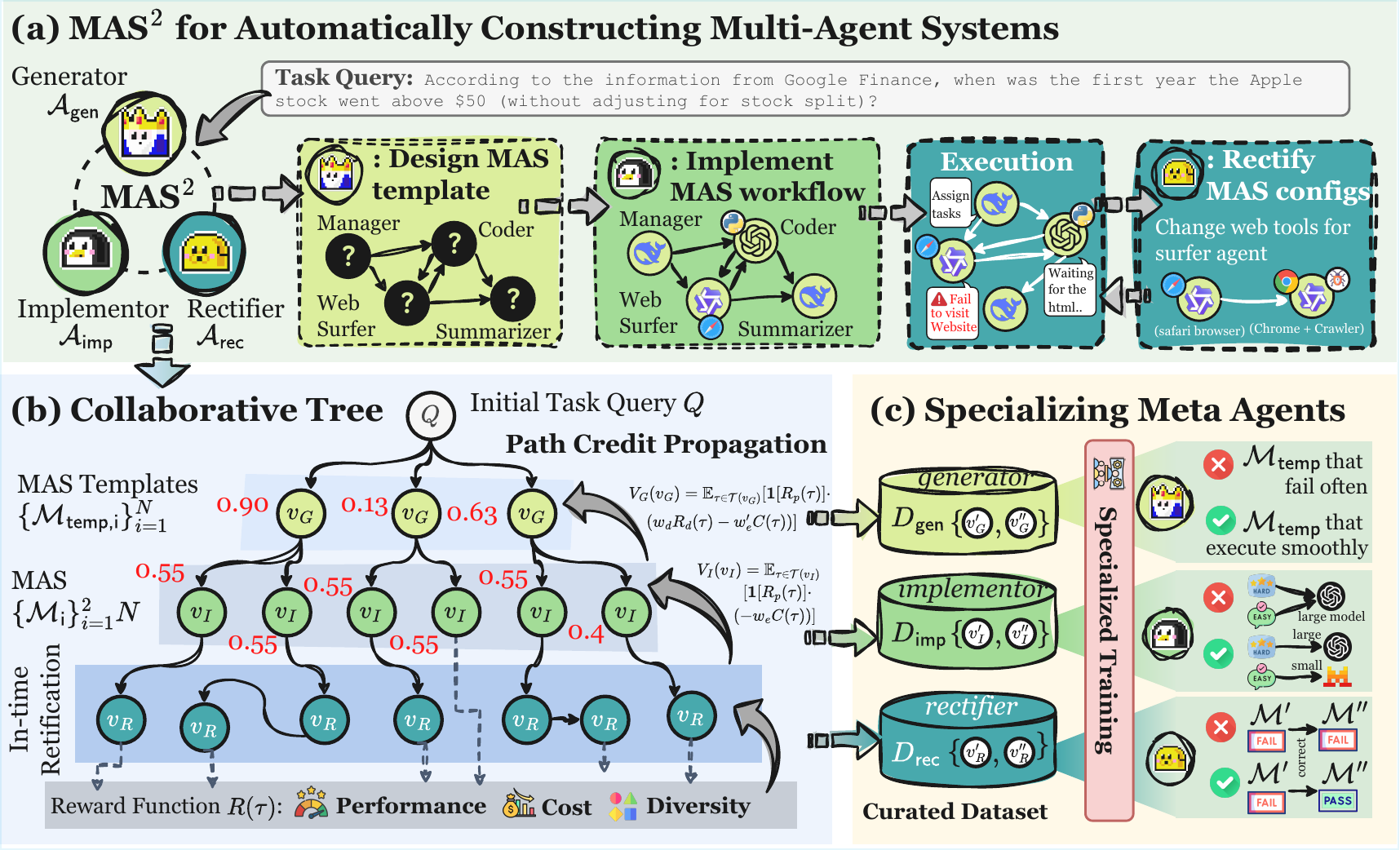}
\vspace{-1.5em}
\caption{The framework overview of our proposed \ourmethod.}
\label{fig:framework}
\vspace{-1.3em}
\end{figure*}

\vspace{-0.3em}
\section{Methodology}
\vspace{-0.4em}
\Cref{fig:framework} presents the overall workflow of \ourmethod. At inference time, \ourmethod accepts each task query as input: the \textit{generator} agent produces a MAS template, the \textit{implementer} instantiates its components, and the \textit{rectifier} continuously monitors and adapts execution in real time ($\triangleright$ \Cref{sec:construct_mas}). These meta-agents are trained under our \textit{collaborative tree optimization} (CTO) framework, which collects preference signals through path-level credit propagation ($\triangleright$ \Cref{sec:curate_data}), and subsequently leverages them for targeted meta-agent training ($\triangleright$ \Cref{sec:train-agent}).

\subsection{Meta-Scheduling of Multi-Agent Systems}
\label{sec:construct_mas}
\vspace{-0.4em}
The effective deployment of an MAS for a novel, complex task necessitates a departure from static, pre-defined architectures. Real-world problems are inherently dynamic, demanding a framework capable of composing and adapting an MAS on-the-fly. Therefore, we introduce our \ourmethod to dynamically construct a bespoke MAS tailored to the specific demands of a given task query, $Q$. 
We begin by formalizing the constitution of a target MAS $\mathcal{M}$ as four core operational components:
\begin{equation}
    \mathcal{M} = \langle\mathcal{R}, \mathcal{P}, \mathcal{T}, \mathcal{B}\rangle,
\end{equation}
where $\mathcal{R} = \{r_1, \dots, r_{|\mathcal{R}|}\}$ denotes a set of allocated agents, $\mathcal{P} = \{\rho_{ij} \mid r_i,r_j \in \mathcal{R}\}$ defines the communication protocol by specifying the permissible message structures $\rho_{ij}$ between agents, $\mathcal{T} = \{t_1, \dots, t_{|\mathcal{T}|}\}$ is a set of available tools (\textit{e.g.}, python interpreter, playwright browser), and $\mathcal{B} = \{b_i \mapsto r_i\}_{r_i \in \mathcal{R}}$ assigns concrete LLM backbones $b_i$ to each agent. 

\vspace{-0.4em}
\paragraph{Generator.} The process commences with the {\textit{generator}} agent $\mathcal{A}_\mathsf{gen}$, which acts as the system architect. Given the query $Q$, the generator formulates a strategic blueprint for the MAS by producing a workflow template, $\mathcal{M}_{\text{temp}}$, which abstracts away the final computational resources:
\begin{equation}\label{eq:generator}
    \mathcal{M}_{\text{temp}} = \langle\mathcal{R}, \mathcal{P}, \mathcal{T}\rangle \sim \pi_\mathsf{gen}(\cdot | Q),
\end{equation}
where $\pi_\mathsf{gen}$ is the generator's policy. \Cref{eq:generator} provides a complete yet uninstantiated MAS plan.

\vspace{-0.4em}
\paragraph{Implementor.} Next, the {\textit{implementor}} agent $\mathcal{A}_\mathsf{imp}$ translates the template $\mathcal{M}_{\text{temp}}$ into a fully executable system. It defines an assignment policy, $\phi$, that maps each role to a concrete LLM backbone from a candidate LLM pool $\mathbb{L} = \{b_1, \dots, b_{|\mathbb{L}|}\}$:
\begin{equation}
    \phi: \mathcal{R} \to \mathbb{L}, 
    \quad r_i \mapsto b_{j(i)} \quad \text{where } \phi \sim \pi_\mathsf{imp}(\cdot | \mathcal{M}_{\text{temp}}, \mathbb{L},Q),
\end{equation}
where $\pi_\mathsf{imp}$ is the implementor's policy and $j(i)$ indexes the selected backbone for agent $r_i$. The final MAS is then composed as
\begin{equation}
    \mathcal{M} = \mathcal{M}_{\text{temp}} \oplus \{(r_i, \phi(r_i)) \mid r_i \in \mathcal{R}\},
\end{equation}
where $\oplus$ denotes the composition that integrates the backbone assignments into the workflow template, producing a fully instantiated, executable MAS. We provide illustrative examples of the multi-agent systems generated by \ourmethod in \Cref{app:workflows}.

\vspace{-0.3em}
\paragraph{Rectifier.} With the MAS $\mathcal{M}$ instantiated, it is deployed to begin execution. The {\textit{rectifier}} agent $\mathcal{A}_\mathsf{rec}$ assumes an online monitoring role to ensure system resilience. The agent is invoked if its trigger function $A_R$, which monitors the execution state $s_t$, evaluates to 1:
\begin{equation}
    A_R(s_t) = \mathbbm{1} \left[ C(s_t) > \theta_C \lor O(s_t) = \text{\texttt{Failure}} \right],
\end{equation}
where $\mathbbm{1}[\cdot]$ is the indicator function. The rectifier is triggered if either the cumulative resource consumption $C(s_t)$ (\textit{e.g.}, token count, execution steps) exceeds a budget $\theta_C$, or if the operational outcome $O(s_t)$ results in an explicit failure (\textit{e.g.}, search engine failure, code execution error). Upon activation, the rectifier intervenes by generating a modification to the current system configuration, $\mathcal{M}_t$. This can range from local adjustments (\textit{e.g.}, re-assign proper tools, modify agent prompts) to global architectural changes (\textit{e.g.}, revise workflow codes). The rectifier's intervention results in:
\begin{equation}
    \mathcal{M}_{t+1} \sim \pi_\mathsf{rec}(\cdot | \mathcal{M}_t, s_t),
\end{equation}
where $\pi_\mathsf{rec}$ is the rectifier's learned policy. The system then resumes execution from state $s_t$ using the updated configuration $\mathcal{M}_{t+1}$. We give illustrative examples of how the rectifier modifies an erroneous MAS in \Cref{app:rectification}. This synergistic process ensures that the generated MAS is not only custom-built for the task but also possesses the capacity for real-time self-correction, a key advantage for tackling long-horizon, unpredictable problem domains.

\vspace{-0.4em}
\subsection{Training Trajectory Curation}
\label{sec:curate_data}
\vspace{-0.6em}
To enable the specialization of our meta-agents, we introduce our \textbf{Collaborative Tree Optimization (CTO)} framework to construct a collaborative decision tree, which serves as the formal basis for our data curation. Formally, we represent the structure as a rooted, directed tree $\mathcal{G}_Q = (V, E)$ associated with a task query $Q$. The vertex set $V$ contains the query root node $v_Q$, a layer of generator nodes $\{v_G\}$, a layer of implementor nodes $\{v_I\}$, rectifier nodes $\{v_R\}$ that may appear adaptively during execution, and terminal leaf nodes $\{v_F\}$. The tree is populated through a sequential sampling process: the generator branches out $K$ candidate templates from the root, the implementor expands each with $N$ executable instantiations, and the rectifier may introduce further branches by adjusting the current MAS. A trajectory $\tau$ is defined as a unique path from the root $v_Q$ to a terminal node $v_F$:
\begin{equation}
    \tau = (v_Q \xrightarrow{\mathcal{A}_\mathsf{gen}} v_G \xrightarrow{\mathcal{A}_\mathsf{imp}} v_I \dots \xrightarrow{\mathcal{A}_\mathsf{rec}} v_F).
\end{equation}

To evaluate such trajectories, a performance-only metric is inadequate, as it ignores the resource efficiency of the constructed MAS. We therefore introduce a conditional, cost-sensitive reward function $R(\tau)$, which assigns zero reward to failed trajectories and evaluates successful ones by their normalized resource consumption:  
\begin{equation}
    R(\tau) = \mathbbm{1}[R_p(\tau)] \cdot \frac{1}{C_{\text{norm}}(\tau)},  \;   C_{\text{norm}}(\tau) = \frac{C(\tau)}{\frac{1}{|\mathcal{T}|}\sum_{\tau' \in \mathcal{T}} C(\tau')},
\end{equation}
where $R_p(\tau)\in\{0,1\}$ denotes the success status of trajectory $\tau$, and $C(\tau)$ measures raw resource consumption (\textit{i.e.}, LLM API cost) and the denominator rescales costs relative to the empirical mean across the trajectory set $\mathcal{T}$. 
 After assigning terminal rewards to the leaves of $\mathcal{G}_Q$, we propagate credit backward to intermediate decision nodes via \textbf{path credit propagation}. The value of a node $v \in V$ is defined as the expected reward of all trajectories that pass through it:
\begin{equation}
    V(v) = \mathbb{E}_{\tau \in \mathcal{T}(v)}[R(\tau)] \approx \frac{1}{|\mathcal{T}(v)|} \sum_{\tau \in \mathcal{T}(v)} R(\tau),
\end{equation}
where $\mathcal{T}(v)$ denotes the set of trajectories traversing $v$. This Monte Carlo estimate attributes terminal outcomes to upstream decisions, thereby providing each node with a principled value signal. The resulting annotated tree $\mathcal{G}_Q$ forms a structured dataset of preference-labeled decisions, which is subsequently used to train our meta-agents.

\vspace{-0.3em}
\subsection{Training and Specializing Meta Agents}
\label{sec:train-agent}
\vspace{-0.3em}
With the collaborative decision tree $\mathcal{G}_Q$ fully annotated with value estimates from the above curation process, the final objective is to specialize the meta-agent policies. A naive preference alignment treating all winning actions equally would discard the rich, quantitative information captured by our node values. To leverage this, following previous relative reward-based preference methods~\citep{gao2024rebelreinforcementlearningregressing,wang2025scoreflowmasteringllmagent}, we leverage a value-guided optimization process to guide the training of the meta agents within \ourmethod.

\paragraph{Value-Guided Preference Construction.}
First, we translate the value-annotated tree into preference data. For each non-terminal node $v \in \mathcal{G}_Q$ representing a decision context $c_v$, we form preference tuples. Unlike standard binary preference pairs, our tuples incorporate the quantitative margin of victory:
\begin{equation}
    \mathcal{D}_{\pi} = \left\{ (c_v, a_{\text{win}}, a_{\text{lose}}, \Delta V) \middle| \begin{aligned} 
    & (v, a_{\text{win}}), (v, a_{\text{lose}}) \in E, \\
    & \Delta V = V(v') - V(v'') > 0
    \end{aligned} \right\}
\end{equation}
where $v'$ and $v''$ are the successor nodes of $a_{\text{win}}$ and $a_{\text{lose}}$ respectively. The term $\Delta V$ represents the ``preference strength," a crucial signal indicating how much better one decision was than another.

\vspace{-0.4em}
\paragraph{Policy Specialization via Value-Scaled Optimization.}
Our training process refines a reference policy $\pi_{\text{ref}}$ (the vanilla LLM) into a specialized policy $\pi^*$ (a specialized generator/implementer/rectifier). Our loss function is designed to incorporate the preference strength $\Delta V$ to modulate the gradient during training. This ensures that the model learns most from high-confidence preference pairs where the value difference is significant, while treating low-margin pairs as less influential. The optimization objective for each policy $\pi_\theta$ is to minimize the following value-scaled loss:
\begin{equation}
    \mathcal{L}_{\text{CTO}}(\pi_\theta; \pi_{\text{ref}}) = - \mathbb{E}_{(c, a_{\text{win}}, a_{\text{lose}}, \Delta V) \sim \mathcal{D}_{\pi}} \left[ \Delta V \cdot \log \sigma\left(\beta \log\frac{\pi_\theta(a_{\text{win}}|c)}{\pi_{\text{ref}}(a_{\text{win}}|c)} - \beta \log\frac{\pi_\theta(a_{\text{lose}}|c)}{\pi_{\text{ref}}(a_{\text{lose}}|c)}\right) \right]
\end{equation}
where $\sigma(\cdot)$ is the logistic function and $\beta$ is a temperature parameter. By weighting each term in the expectation by $\Delta V$, our loss function directly prioritizes learning from the most unambiguous and impactful decisions identified during trajectory curation.

This optimization procedure is applied independently to the generator, implementor, and rectifier, yielding three specialized policies, $\pi^*_\mathsf{gen}$, $\pi^*_\mathsf{imp}$, and $\pi^*_\mathsf{rec}$.  Together, these components establish a unified methodology that not only enables the dynamic construction of bespoke multi-agent systems, but also ensures their continual refinement through curated training and value-guided specialization.

\section{Experiments}
\vspace{-0.5em}
\subsection{Experimental Setup}\label{sec:setup}
\vspace{-0.5em}
\paragraph{Benchmarks.} To provide a comprehensive evaluation of \ourmethod, we employ eight benchmarks spanning four domains: \ding{110} \textbf{multi-hop search}, including HotpotQA~\citep{yang2018hotpotqadatasetdiverseexplainable}, Bamboogle~\citep{press2022measuring}, and Natural Question (NQ)~\citep{kwiatkowski-etal-2019-natural}; \ding{110} \textbf{deep research}, represented by BrowseComp+~\citep{chen2025browsecomp-plus}, an enhanced variant of OpenAI's BrowseComp~\citep{wei2025browsecompsimplechallengingbenchmark} that corrects erroneous cases and enables more stable offline evaluation; \ding{110} \textbf{code generation}, comprising HumanEval~\citep{chen2021evaluatinglargelanguagemodels} and MBPP~\citep{austin2021programsynthesislargelanguage}; and \ding{110} \textbf{mathematical reasoning}, assessed with MATH~\citep{hendrycks2021measuringmassivemultitasklanguage}.

\vspace{-0.8em}
\paragraph{Baselines.} The baselines against which we compare \ourmethod\ can be broadly grouped into three categories: \ding{110} \textbf{vanilla LLMs}, where single models are directly evaluated, alongside prompting strategies such as CoT~\citep{wei2023cot} and Self-Consistency~\citep{wang2023selfconsistencyimproveschainthought}; \ding{110} \textbf{handcrafted multi-agent systems}, including MedPrompt~\citep{nori2023medpropmt}, MultiPersona~\citep{wang2023multipersona}, LLM-Debate~\citep{du2023improvingfactualityreasoninglanguage}, and DyLAN~\citep{liu2024dynamicllmpoweredagentnetwork}; \ding{110} \textbf{(partially) automated multi-agent systems}, comprising ADAS~\citep{hu2025adas}, MaAS~\citep{zhang2025multiagentarchitecturesearchagentic}, AFlow~\citep{zhang2024aflow}, and ScoreFlow~\citep{wang2025scoreflowmasteringllmagent}. For multi-agent system baselines, we instantiate them using the relatively highest-performing QwQ-32B and GPT-4o and report the average scores.

\vspace{-0.8em}
\paragraph{Parameter \& Model Configurations.} The LLM pool available for \ourmethod is specified as GPT-4o and GPT-4o-mini~\citep{liu2024dynamicllmpoweredagentnetwork}, Qwen2.5-72b-instruct~\citep{qwen2025qwen25technicalreport}, Qwen3-14b~\citep{yang2025qwen3}, and QwQ-32B~\citep{qwenlmQwQ32BEmbracing}. We access the above models via the OpenRouter~\footnote{\url{https://openrouter.ai/}} API. The backbone LLM for the generator, implementer, and rectifier in \ourmethod are consistently set as Qwen3-8B~\citep{yang2025qwen3}. For data curation in \Cref{sec:curate_data}, the generator expands into four candidate $v_G$ nodes, and each $v_G$ is further instantiated twice by the implementor, yielding two corresponding $v_I$ nodes per branch. In the RL training in \Cref{sec:train-agent}, we set \texttt{epoch\_num} $=2$, \texttt{learning\_rate} $=$ \num{5e-5}, and \texttt{gradient\_accumulation\_steps} $=4$. We finetune the base Qwen3-8B via LoRA, with \texttt{rank} $=$ \num{8} and $\texttt{alpha}=16$.

\begin{table*}[!tbp]
\centering
\small
\caption{
\textbf{Performance comparison} across 8 benchmarks across 13 baselines. Each cell reports the average of \textbf{three random runs}. The best/second-best results are \textbf{bolded}/\underline{underlined}.} 
\vspace{-0.9em}
\setlength{\tabcolsep}{1.0pt}
\renewcommand{\arraystretch}{1.2}
\label{tab:main_results}
\begin{tabular}{l ccc c ccc c}
 \Xhline{1.2pt}
 & \multicolumn{3}{c}{\textbf{Multi-hop Search}} & \multicolumn{1}{c}{\textbf{Deep Research}} & \multicolumn{2}{c}{\textbf{Code Generation}} & \multicolumn{1}{c}{\textbf{Math}} \\
\cmidrule(lr){2-4} \cmidrule(lr){5-5} \cmidrule(lr){6-7} \cmidrule(lr){8-8}
\multirow{-2}{*}{\textbf{Model}} & HotpotQA & Bamboogle & NQ & BrowseComp+ & HumanEval & MBPP & MATH \\
 \Xhline{0.5pt}
\rowcolor{CadetBlue!10}Qwen3-14B & $62.2$ & $34.4$ & $51.9$ & $13.1$ & $75.6$ & $40.8$ & $61.5$ \\
GPT-4o-mini & $64.1$ & $26.4$ & $70.7$ & $13.7$ & $87.8$ & $46.0$ & $51.2$ \\
\rowcolor{CadetBlue!10}QwQ-32B & $64.9$ & $39.2$ & $61.5$ & $2.8$ & $51.8$ & $82.2$ & $65.6$ \\
Qwen-2.5-72B & $66.8$ & $31.2$ & $63.0$ & $2.6$ & $82.3$ & $76.0$ & $55.3$ \\
\rowcolor{CadetBlue!10}GPT-4o & $69.5$ & $49.6$ & $71.1$ & $13.2$ & $89.6$ & $73.4$ & $56.5$ \\
 \Xhline{0.5pt}
Avg. (Above) & $65.5$ & $36.2$ & $63.6$ & $9.5$ &$ 77.4$ & $63.7$ &$ 58.0$ \\
 \Xhline{0.5pt}
CoT (GPT-4o)  & $66.2_{\color{ForestGreen}{\uparrow 0.7}}$ & $52.8_{\color{ForestGreen}{\uparrow 16.6}}$ & $68.7_{\color{ForestGreen}{\uparrow 5.1}}$ & \uline{$16.1_{\color{ForestGreen}{\uparrow 7.6}}$} & $90.8_{\color{ForestGreen}{\uparrow 13.4}}$ & $75.4_{\color{ForestGreen}{\uparrow 11.7}}$ & $58.9_{\color{ForestGreen}{\uparrow 0.9}}$ \\
\rowcolor{CadetBlue!10}SC (GPT-4o) & $66.4_{\color{ForestGreen}{\uparrow 0.9}}$ & $54.4_{\color{ForestGreen}{\uparrow 18.2}}$ & $73.6_{\color{ForestGreen}{\uparrow 10.0}}$ & $11.8_{\color{ForestGreen}{\uparrow 2.3}}$ & \uline{$96.3_{\color{ForestGreen}{\uparrow 18.9}}$} & $75.1_{\color{ForestGreen}{\uparrow 11.4}}$ & ${62.0_{\color{ForestGreen}{\uparrow 4.0}}}$ \\
  \Xhline{0.5pt}
{MedPrompt} & $72.4_{\color{ForestGreen}{\uparrow 6.9}}$ & $48.0_{\color{ForestGreen}{\uparrow 11.8}}$ & $66.1_{\color{ForestGreen}{\uparrow 2.5}}$ & $12.0_{\color{ForestGreen}{\uparrow 2.5}}$ & $92.1_{\color{ForestGreen}{\uparrow 14.7}}$ & $69.2_{\color{ForestGreen}{\uparrow 5.5}}$ & $59.7_{\color{ForestGreen}{\uparrow 1.7}}$ \\
\rowcolor{CadetBlue!10} {MultiPersona} & $71.1_{\color{ForestGreen}{\uparrow 5.6}}$ & $50.2_{\color{ForestGreen}{\uparrow 14.0}}$& $68.4_{\color{ForestGreen}{\uparrow 4.8}}$ & $10.1_{\color{ForestGreen}{\uparrow 0.6}}$ & $92.9_{\color{ForestGreen}{\uparrow 15.5}}$ & $70.4_{\color{ForestGreen}{\uparrow 6.7}}$ & $53.7_{\color{red}{\downarrow 4.3}}$ \\
{LLM-Debate} & $66.9_{\color{ForestGreen}{\uparrow 1.4}}$ & $54.4_{\color{ForestGreen}{\uparrow 18.2}}$ & $70.8_{\color{ForestGreen}{\uparrow 7.2}}$ & $15.3_{\color{ForestGreen}{\uparrow 5.8}}$ & $88.7_{\color{ForestGreen}{\uparrow 11.3}}$ & $70.3_{\color{ForestGreen}{\uparrow 6.6}}$ & $67.3_{\color{ForestGreen}{\uparrow 9.3}}$ \\
\rowcolor{CadetBlue!10}{DyLAN} & $80.8_{\color{ForestGreen}{\uparrow 15.3}}$ & $59.7_{\color{ForestGreen}{\uparrow 23.5}}$ & $72.1_{\color{ForestGreen}{\uparrow 8.5}}$ & $15.8_{\color{ForestGreen}{\uparrow 6.3}}$ & $90.4_{\color{ForestGreen}{\uparrow 13.0}}$ & $77.3_{\color{ForestGreen}{\uparrow 13.6}}$ & $65.7_{\color{ForestGreen}{\uparrow 7.7}}$ \\
 \Xhline{0.5pt}
{ADAS} & $78.5_{\color{ForestGreen}{\uparrow 13.0}}$ & $50.8_{\color{ForestGreen}{\uparrow 14.6}}$  & $65.9_{\color{ForestGreen}{\uparrow 2.3}}$ & $7.0_{\color{red}{\downarrow 2.5}}$ & $88.8_{\color{ForestGreen}{\uparrow 11.4}}$ & $68.7_{\color{ForestGreen}{\uparrow 5.0}}$ & $51.7_{\color{red}{\downarrow 6.3}}$ \\
\rowcolor{CadetBlue!10}{MaAS} & $83.6_{\color{ForestGreen}{\uparrow 18.1}}$ & $62.0_{\color{ForestGreen}{\uparrow 25.8}}$ & $76.0_{\color{ForestGreen}{\uparrow 12.4}}$ & $14.0_{\color{ForestGreen}{\uparrow 4.5}}$ & $92.8_{\color{ForestGreen}{\uparrow 15.4}}$ & $82.2_{\color{ForestGreen}{\uparrow 18.5}}$ & $70.1_{\color{ForestGreen}{\uparrow 12.1}}$ \\
{AFlow} & $77.9_{\color{ForestGreen}{\uparrow 12.4}}$ & $59.2_{\color{ForestGreen}{\uparrow 23.0}}$ &  $74.5_{\color{ForestGreen}{\uparrow 10.9}}$& $10.0_{\color{ForestGreen}{\uparrow 0.5}}$& $92.9_{\color{ForestGreen}{\uparrow 15.5}}$ & $82.9_{\color{ForestGreen}{\uparrow 19.2}}$ & $68.5_{\color{ForestGreen}{\uparrow 10.5}}$ \\
\rowcolor{CadetBlue!10}{ScoreFlow} & \uline{$86.0_{\color{ForestGreen}{\uparrow 20.5}}$} & \uline{$64.8_{\color{ForestGreen}{\uparrow 28.6}}$} & \uline{$76.4_{\color{ForestGreen}{\uparrow 12.8}}$} & $10.4_{\color{ForestGreen}{\uparrow 0.9}}$ & $95.9_{\color{ForestGreen}{\uparrow 18.5}}$ & \uline{$84.7_{\color{ForestGreen}{\uparrow 21.0}}$} & $64.4_{\color{ForestGreen}{\uparrow 6.4}}$ \\
 \Xhline{0.5pt}
\ourmethod & $\bm{89.3_{\color{ForestGreen}{\uparrow 23.8}}}$ & $\bm{67.2_{\color{ForestGreen}{\uparrow 31.0}}}$ & $\bm{79.1_{\color{ForestGreen}{\uparrow 15.5}}}$ &  $\bm{19.7_{\color{ForestGreen}{\uparrow10.2}}}$ & $\bm{97.0_{\color{ForestGreen}{\uparrow 19.6}}}$ & $\bm{85.1_{\color{ForestGreen}{\uparrow 21.4}}}$ & \bm{$71.3_{\color{ForestGreen}{\uparrow 13.3}}$} \\
 \Xhline{1.2pt}
 \vspace{-1.3em}
\end{tabular}
\vspace{-1.5em}
\end{table*}

\subsection{Main Results}
\vspace{-0.7em}

\Cref{tab:main_results} compares \ourmethod against all single-LLM baselines within the LLM pool, as well as eight representative MAS baselines, from which we distill the following key observations:

\paragraph{Obs. \ding{182}: Existing automated MAS fail to generalize across domains.}

\vspace{-0.5em}
\Cref{tab:main_results} shows that current MAS baselines perform inconsistently across tasks. For example, {MultiPersona} performs well on HumanEval ($92.9\%$) and Bamboogle ($50.2\%$) yet drops on MATH ($-4.3\%$) compared with single LLMs. {LLM-Debate} excels on Math ($67.3\%$) and Bamboogle ($54.4\%$) but lacks BrowseComp+ coverage. Stronger baselines like {ADAS} and {MaAS} show mixed performance: {ADAS} scores $78.5\%$ on HotpotQA and $88.8$ on HumanEval but falls on deep research (merely $7.0\%$ on BrowseComp+) and MATH ($51.7\%$), while {MaAS} is strong on multi-hop search ($83.6\%$ on HotpotQA) and code generation ($92.8\%$ on HumanEval) yet remains weak on BrowseComp+ ($14.0\%$). Overall, these results indicate that existing MASs often excel in some domains but fail to generalize consistently.

\vspace{-0.5em}
\paragraph{Obs. \ding{183}:  \ourmethod outperforms both handcrafted and automated multi-agent systems.}
\textbf{(I)} Compared to handcrafted MAS, \ourmethod obtains SOTA performance across multiple domains, such as mathematics, code, and deep research. Concretely, \ourmethod achieves $89.3\%$ on HotpotQA, surpassing the best manual setting DyLAN by $8.5\%$, and outperforming MedPrompt by $16.9\%$. In MATH, \ourmethod scores $71.3\%$, slightly outperforming LLM-Debate by $4.0\%$.  
\textbf{(II)} \ourmethod also excels when compared to automated MAS. In the multi-hop search tasks (HotpotQA, Bamboogle, NQ), \ourmethod achieves an average of $\sim78.5\%$, surpassing ScoreFlow and MaAS by $2.8\%$ and $4.7\%$, respectively. In code generation, \ourmethod scores $97.0\%$ on HumanEval, outperforming AFlow by $4.1\%$. In the MATH task, \ourmethod achieves $71.3\%$, improving by $6.9\%$ over ScoreFlow and $2.8\%$ over AFlow. In summary, \ourmethod consistently demonstrates strong performance across multiple task domains, validating its robust multi-task adaptability and effectiveness.

\vspace{-0.5em}
\subsection{Genealization Study}
\vspace{-0.3em}

Although \ourmethod was trained using a fixed LLM pool, including models such as GPT-4o, we aim to demonstrate that its capabilities extend beyond previously seen LLMs and can effectively leverage unseen models. To evaluate this, we augment the LLM pool at inference with three stronger, previously unseen LLMs (Qwen3-Coder~\citep{qwen3-coder}, GPT-5-Mini, and Gemini-2.5-Pro~\citep{comanici2025gemini}) while retaining the original models.

\vspace{-0.5em}
\paragraph{Obs. \ding{184}: \ourmethod generalizes to and exploits previously unseen LLM backbones.}
\begin{wraptable}{r}{0.57\linewidth}
\centering
\scriptsize
\caption{Performance results when integrated with stronger unseen LLMs. ``Vanilla LLM'' denotes merely using one LLM backbone for testing. Checkmarks are used to denote the LLMs used.}
\vspace{-0.9em}
\setlength{\tabcolsep}{2pt}
\renewcommand{\arraystretch}{1.2}
\label{tab:reduced_results}
\begin{tabular}{l ccc | cc | cc}
\Xhline{1.2pt}
\multicolumn{1}{l}{} & \multicolumn{3}{c}{\textbf{Stronger LLMs}} & \multicolumn{4}{c}{\textbf{Datasets}} \\
\cmidrule(lr){2-4} \cmidrule(lr){5-8}
\textbf{Method} & \multirow{2}{*}{\makecell{Qwen3-\\Coder}} & \multirow{2}{*}{\makecell{GPT-5-\\Mini}} & \multirow{2}{*}{\makecell{Gemini-\\2.5-Pro}} & \multicolumn{2}{c}{MATH} & \multicolumn{2}{c}{Bamboogle} \\
\cmidrule(lr){5-6} \cmidrule(lr){7-8}
& & & & Perf.(\%) & Cost(\$) & Perf.(\%) & Cost(\$) \\
\Xhline{0.5pt}
\multirow{3}{*}{\makecell{Vanilla\\LLM}}
& {\color{teal}\CheckmarkBold} & & & 69.7 & 0.54 & 32.8 & 0.06 \\
& & & {\color{teal}\CheckmarkBold} & 68.0 & 14.48 & 83.2 & 2.27\\
& & {\color{teal}\CheckmarkBold} & & 87.3 & 1.22 & 80.8 &  0.21\\
\midrule
\multirow{4}{*}{\ourmethod}
&{\color{red}\XSolidBrush}  &  {\color{red}\XSolidBrush} & {\color{red}\XSolidBrush} & 71.3 & 0.74 & 67.2 &  0.15\\
& {\color{teal}\CheckmarkBold} & {\color{teal}\CheckmarkBold}  & {\color{red}\XSolidBrush}  & 88.6 & 2.22 & 82.5 & 0.53 \\
& {\color{teal}\CheckmarkBold} & {\color{red}\XSolidBrush} & {\color{teal}\CheckmarkBold} & 84.8 & 15.47 & 84.2 & 2.57 \\
& {\color{red}\XSolidBrush} & {\color{teal}\CheckmarkBold} & {\color{teal}\CheckmarkBold} & 90.6 & 16.14 & 84.0 &2.73\\
\Xhline{1.2pt}
\vspace{-1.7em}
\end{tabular}
\end{wraptable}

\Cref{tab:reduced_results} shows that \ourmethod effectively integrates these unseen LLM backbones, significantly improving performance. For instance, using the vanilla Qwen3-Coder, performance on MATH is $69.7\%$ (cost $\$0.54$) and $32.8\%$ on Bamboogle (cost $\$0.06$). When \ourmethod integrates Qwen3-Coder and GPT-5-Mini, MATH performance improves to $88.6\%$ (cost $\$2.22$) and Bamboogle to $82.5\%$ (cost $\$0.53$). Further integration with GPT-5-Mini and Gemini-2.5-Pro boosts MATH performance to $90.6\%$ (cost $\$16.14$) and Bamboogle to $84.0\%$ (cost $\$2.73$), offering significant gains over the vanilla LLM baseline while maintaining a reasonable cost increase. These results confirm that \ourmethod not only generalizes to new LLM backbones but also harnesses their strengths to achieve higher accuracy, without requiring additional fine-tuning. This validates the flexibility of \ourmethod for practical deployment.

\vspace{-0.5em}
\subsection{Cost Analysis}
\vspace{-0.5em}

\begin{wrapfigure}{R}{0.7\textwidth} 
\centering
\vspace{-5mm}
\includegraphics[width=0.69\textwidth]{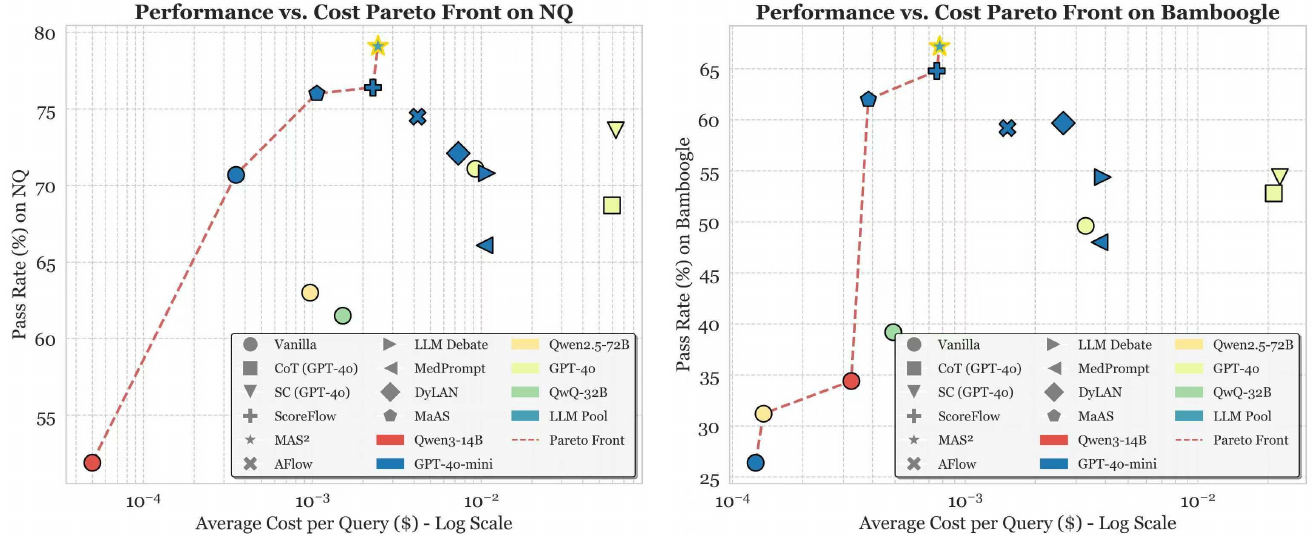}
\caption{\small Cost-performance trade-off on the NQ and Bamboogle benchmark.} \label{fig:cost}
\vspace{-3.5mm}
\end{wrapfigure}

To evaluate the practical viability of our method, we analyze its economic efficiency by exploring the performance-cost trade-off. As shown in \Cref{fig:cost}, we compare \ourmethod against a wide range of baselines to demonstrate its ability to achieve superior results without incurring prohibitive costs.
\vspace{-3.5mm}

\paragraph{Obs. \ding{185}: \ourmethod establishes a new cost-performance Pareto frontier.}

\Cref{fig:cost} demonstrates that \ourmethod (indicated by the \ding{72} symbol) consistently achieves a state-of-the-art balance, delivering higher accuracy than methods of similar cost (\textit{e.g.}, ScoreFlow), and significantly lower cost than methods with comparable accuracy. For instance, on Bamboogle, \ourmethod achieves a $12.8\%$ higher pass rate than the expensive SC (GPT-4o) baseline while being over $25\times$ cheaper. Similarly, on NQ, it surpasses the strong ScoreFlow baseline by $2.7\%$ in accuracy at a nearly identical cost, while simultaneously outperforming SC (GPT-4o) by $5.5$ points and cutting costs by over $20$ times. This Pareto efficiency mainly stems from \ourmethod's ability to operate dynamically at (1) the LLM level, assigning simpler tasks to smaller LLMs and reserving larger, more capable LLMs for complex reasoning, and also (2) system-level, configuring lightweight multi-agent systems for easy problems while deploying more sophisticated system architectures for challenging ones, achieving both accuracy and cost-effectiveness.

\subsection{Framework Analysis}

\begin{figure}
\centering
\vspace{-3em}
\includegraphics[width=0.6\textwidth]{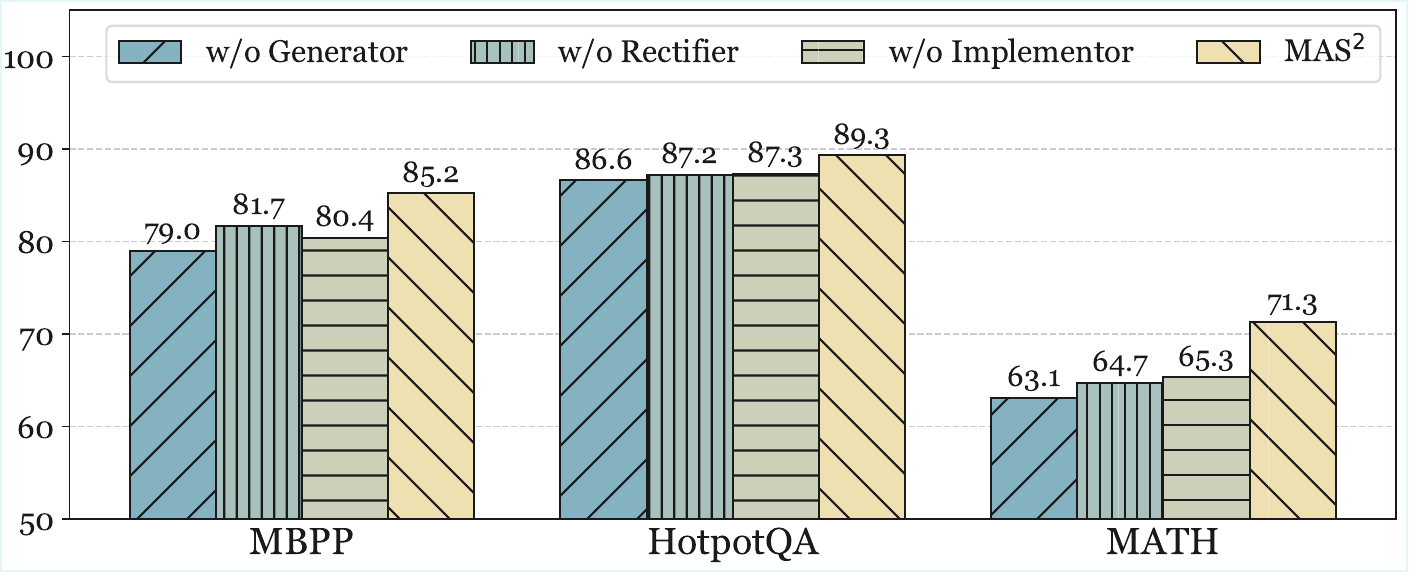}
\vspace{-.8em}
\caption{The ablation study of \ourmethod.}
\label{fig:ablation}
\vspace{-1.2em}
\end{figure}

\paragraph{Ablation Study.}
We conduct ablation studies on MBPP, HotpotQA, and MATH to assess the contribution of each \ourmethod component. Specifically, we consider three ablations: \textit{w/o Generator} and \textit{w/o Implementor} correspond to using the untrained Qwen3-8B model as the generator and implementor, respectively, while \textit{w/o Rectifier} denotes the removal of the rectifier, which monitors and modifies the MAS dynamically. As shown in \Cref{fig:ablation}, removing the generator (\textit{w/o Generator}) results in substantial performance drops (from $85.2\%$ to $79.0\%$ on MBPP, from $89.3\%$ to $86.6\%$ on HotpotQA). Similarly, \textit{w/o Implementor} also causes notable degradation, with scores of $80.4\%$ on MBPP, $87.3\%$ on HotpotQA, and $65.3\%$ on MATH, demonstrating that the trained implementor agent effectively learns to allocate LLMs optimally for different tasks.  \textit{w/o Rectifier} leads to intermediate drops of $3.5\%$ on MBPP and $6.6\%$ on MATH, indicating that the rectifier is crucial for real-time self-adjustment during interactions with the environment. These results confirm that all three modules are indispensable, and omitting any component leads to clear performance degradation across multiple domains.

 \vspace{-1em}

\begin{figure}[!h]
\centering
\vspace{-0.4em}
\includegraphics[width=1.0\linewidth]{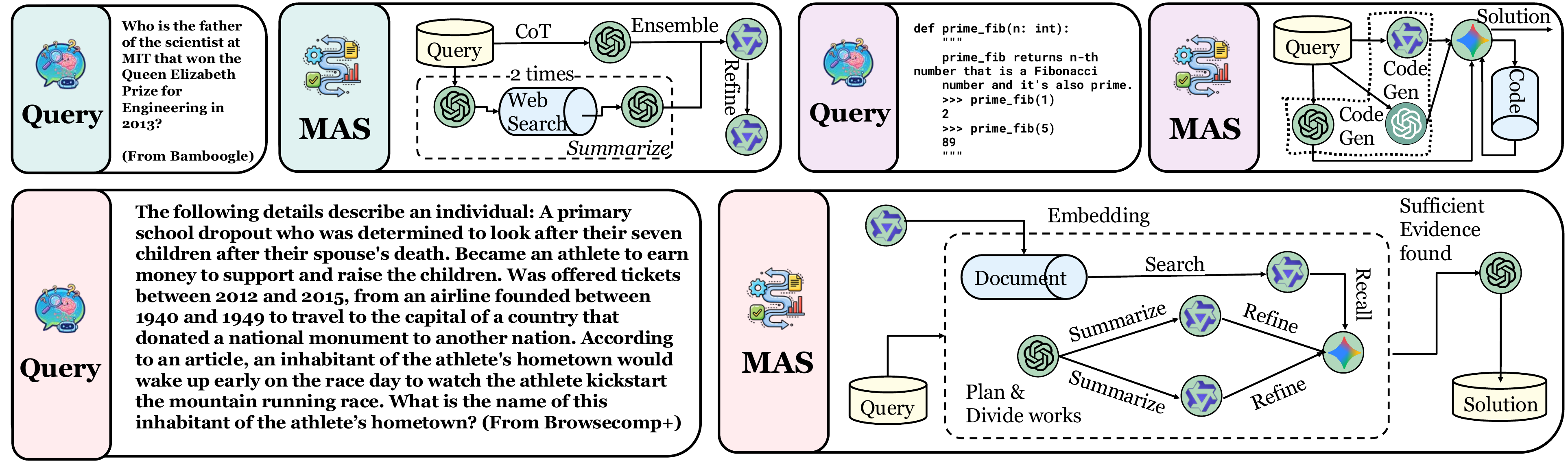}
\vspace{-1.6em}
\caption{A visualization of task-specific multi-agent systems designed by \ourmethod on Bamboogle, HumanEval, and BrowseComp+ benchmarks.}
\label{fig:case_study}
\vspace{-1.em}
\end{figure}

\paragraph{Case Study}
\Cref{fig:case_study} shows task-specific multi-agent systems designed by \ourmethod that strategically allocate different LLMs, each with customized complexity and workflow configurations. 

The explanation of each devised multi-agent system is as follows:
\begin{itemize}[leftmargin=*,itemsep=0pt]
\item \textbf{(I)} \textbf{Bamboogle (QA)}: In QA tasks, the designed MAS utilizes multiple models (e.g., Qwen series) to generate candidate answers from different perspectives. Subsequently, a premier model (e.g., Gemini-2.5-pro) takes on the critical role of evaluating these candidates to deliver a final verdict.

\item \textbf{(II)} \textbf{HumanEval (Coding)}: For code generation, the MAS design embodies a division of labor. More economical, lightweight models are assigned to generate and test multiple code solutions in parallel. Finally, a top-tier model (e.g., GPT-4o) is responsible for reviewing these initial proposals and synthesizing a final, high-quality program.

\item \textbf{(III)} \textbf{BrowseComp+ (Research)}: In the MAS designed for complex research tasks, cognitive roles are clearly demarcated. The most powerful model (e.g., GPT-4o) focuses on high-level cognitive activities such as reflective reasoning, strategic planning, and information synthesis. Meanwhile, more cost-effective models are employed to execute the mechanical, high-throughput tasks of searching and retrieving information.
\end{itemize}

Results showcase how \ourmethod flexibly designs heterogeneous MAS tailored to domain-specific demands. In the QA scenario, various models generate diverse initial answers, while a flagship model like GPT-4o is reserved for the final, decisive judgment. In contrast, for code generation, economical models handle parallel generation and testing, with the results ultimately being refined and elevated by a more powerful model. In the research tasks, \ourmethod dynamically coordinates models of different tiers to handle retrieval, summarization, and reflective reasoning. It is noteworthy that the MAS presented here are generated solely by the generator and implementor agents. In \Cref{app:rectification}, we further illustrate potential errors encountered by these MAS and demonstrate how the rectifier dynamically corrects them, thereby validating the adaptive capabilities of \ourmethod.


\vspace{-0.4em}
\section{Conclusion}  
\vspace{-0.6em}
This work advances the frontier of automatic MAS by introducing \ourmethod, a paradigm grounded in recursive self-generation. Unlike prevailing approaches that follow a static ``generate-once-and-deploy'' doctrine, \ourmethod instantiates a meta-level tri-agent architecture, comprising generator, implementer, and rectifier agents, that dynamically composes and adaptively rectifies task-specific MAS in response to real-time demands. Through the proposed Collaborative Tree Optimization, these meta-agents are trained and specialized, enabling robust adaptability and heightened efficiency.  
Empirical studies across seven benchmarks demonstrate that \ourmethod not only achieves consistent gains over state-of-the-art MAS but also scales synergistically with stronger LLM backbones.



\bibliography{iclr2024_conference}
\bibliographystyle{iclr2024_conference}
\clearpage
\appendix

\section{Utilization of Large Language Models}
In the preparation of this manuscript, LLMs are utilized to refine the language and formatting of initial drafts, aid in literature review and data visualization, and facilitate information retrieval.

\section{Prompt Set}
\label{appendix:prompts}

This appendix details the system prompts used to instruct the three core meta-agents within the \ourmethod framework: the Generator, the Implementer, and the Rectifier.

\vspace{1em}

\begin{tcolorbox}[
    notitle, sharp corners, breakable,
    colframe=Periwinkle, colback=white,
    boxrule=2pt, boxsep=1pt, enhanced,
    shadow={2pt}{-2pt}{0pt}{opacity=0.8,mygrey},
    title={\faCogs\quad Prompt for the Generator Agent},
]
{\scriptsize
\textbf{Description:} This prompt instructs the \textbf{Generator} agent. Its objective is to design a high-level, model-agnostic multi-agent system (MAS) workflow template based on a given user query.
\begin{lstlisting}[language=]
"""
You are an expert in designing multi-agent system (MAS) workflows. Your task is to generate a Python script that defines a workflow for solving a given problem. The workflow must be encapsulated within a class named `Workflow`.

## Rules and Constraints:
1.  **Class Structure**: The entire workflow logic must be contained within a single Python class named `Workflow`.
2.  **Output Format**: Your entire response must be enclosed within `<graph>` and `</graph>` XML tags. Do not include any text or explanations outside of these tags.
3.  **Core Logic**: The main execution logic must be within an `async def run_workflow(self)` method inside the `Workflow` class.
4.  **Operators**: You must use the predefined operators listed below to construct the workflow. Do not invent new operators.
5.  **LLM Placeholder**: For any operator that requires an LLM, use the string placeholder `"llm_symbol"`. Do not use specific model names like "gpt-4o".

## Available Operators:
You can use the following operators to build your workflow. They are available under the `operator` module.

-   **`operator.Custom(llm_config, instruction, output_type)`**: A general-purpose operator that takes an instruction and produces text.
-   **`operator.Search(llm_config, query_instruction, context_instruction)`**: An operator designed to perform web searches.
-   **`operator.Programmer(llm_config, instruction)`**: A specialized operator for generating code.
-   **`operator.Reviewer(llm_config, instruction)`**: An operator for reviewing and critiquing code or text.
-   **`operator.ScEnsemble(llm_config, instruction, num_candidates)`**: An operator for generating multiple candidate solutions and selecting the best one (Self-Consistency).
-   **`operator.Tool(tool_name)`**: An operator for executing a specific tool, like a Python interpreter (`tool_name="python"`).

## Example Workflow Structure:
```python
# All necessary imports are provided automatically.

class Workflow:
    def __init__(self, problem: str, **kwargs):
        self.problem = problem
        # Define your operators here
        self.planner = operator.Custom("llm_symbol", instruction="Create a plan.")
        self.coder = operator.Programmer("llm_symbol", instruction="Write the code.")

    async def run_workflow(self):
        # Implement the workflow logic here by calling the operators
        plan = await self.planner(self.problem)
        code = await self.coder(plan)
        return code
Now, based on these instructions, please generate a workflow for the following problem.

Problem:
"""


***

## ## **`END_PROMPT` (End of the Prompt)**

This part provides the final instruction to start generating the code.

```python
"""
<graph>
# Remember to start your Python code here, beginning with the class definition.
class Workflow:
"""
\end{lstlisting}
}
\end{tcolorbox}

\begin{tcolorbox}[
    notitle, sharp corners, breakable,
    colframe=Periwinkle, colback=white,
    boxrule=2pt, boxsep=1pt, enhanced,
    shadow={2pt}{-2pt}{0pt}{opacity=0.8,mygrey},
    title={\faRocket\quad Prompt for the Implementer Agent},
]
{\scriptsize
\textbf{Description:} This prompt is provided to the \textbf{Implementer} agent. It takes the abstract workflow template from the Generator and instantiates it into an executable system by assigning a specific LLM backbone to each agent role from a predefined pool of available models.
\begin{lstlisting}[language=]
"""
You are an expert model selector for AI workflows.

## Goal
Your primary goal is to replace every string placeholder "llm_symbol" in the operator constructors with the most suitable model name for that operator. 

**DO NOT** modify any function signatures or add any extra keyword arguments like `llm=` or `model=` to method calls. 

**ONLY** change the first argument of operator constructors. For example: `operator.Custom("llm_symbol", ...)` should become `operator.Custom("gpt-4o-mini", ...)`.

## Available LLMs
Here are the available LLMs you can choose from:
- **gpt-4o-mini**
- **gpt-4o**
- **qwen/qwen-2.5-72b-instruct**
- **qwen/qwq-32b**

## Reference LLM Catalog
This catalog provides brief descriptions of each model to help you make an informed decision:
- **gpt-4o-mini**: GPT-4o Mini is a smaller, faster variant of OpenAI's GPT-4o multimodal model. It's optimized for lower latency and is best suited for lightweight tasks or when speed is prioritized over peak performance.
- **gpt-4o**: GPT-4o is OpenAI's flagship multimodal model, offering exceptional performance in complex reasoning, including mathematical proofs and multi-step derivations. Its long-context capabilities make it ideal for high-precision evaluations.
- **qwen/qwen-2.5-72b-instruct**: This is a 72-billion-parameter instruction-tuned model that excels in advanced mathematical reasoning, theorem verification, and long-form derivations. It's one of the most powerful open-source models for high-stakes reasoning tasks.
- **qwen/qwq-32b**: This is a medium-sized, reasoning-optimized model from Qwen that is strong at step-by-step multi-hop reasoning and QA.

## Operator Descriptions
Here's what each operator does:
- **Custom**: Generates detailed step-by-step analysis and reasoning for factual questions, potentially using provided context.
- **AnswerGenerate**: Directly generates concise final answers for factual QA tasks based on evidence and reasoning.
- **ScEnsemble**: Evaluates multiple answer candidates and selects the most accurate one for factual questions.
- **Review**: Critiques and refines solutions for factual questions, ensuring accuracy and completeness.

## Critical Instructions
- **Only replace "llm_symbol" strings**; do not change "operator." paths.
- **Keep the exact same operator paths** as in the input code.
- **Do not add any import statements** or module prefixes.
- Your output **must be valid Python code** that can be executed.
- Your final output **must only be code**, starting with `<graph>` and ending with `</graph>`.

## Workflow Code to Modify
<graph>
class Workflow:
    def __init__(self, problem) -> None:
        self.problem = problem
        self.custom = operator.Custom("llm_symbol", self.problem)
        self.sc_ensemble = operator.ScEnsemble("llm_symbol", self.problem)

    async def run_workflow(self):
        """
        This is a workflow graph for the NQ dataset.
        """
        # Analyze the question with context
        analysis = await self.custom(instruction="Can you analyze this question and provide relevant information from the context?")
        # Ensemble multiple reasoning approaches
        final_answer = await self.sc_ensemble(solutions=[analysis])

        return {"solution": final_answer}
</graph>
"""
\end{lstlisting}
}
\end{tcolorbox}

\begin{tcolorbox}[
    notitle, sharp corners, breakable,
    colframe=Periwinkle, colback=white,
    boxrule=2pt, boxsep=1pt, enhanced,
    shadow={2pt}{-2pt}{0pt}{opacity=0.8,mygrey},
    title={\faWrench\quad Prompt for the Rectifier Agent},
]
{\scriptsize
\textbf{Description:} This prompt activates the \textbf{Rectifier} agent when the executing MAS encounters a failure (e.g., code execution error, API timeout) or enters a non-productive state. The agent's task is to analyze the runtime context and error logs to propose a concrete modification to the workflow code.
\begin{lstlisting}[language=]
"""
You are an expert AI workflow debugger. Your task is to analyze and fix a broken Python workflow script based on the provided error log.

## Goal
Identify the root cause of the error in the "Broken Workflow" and correct the code. The corrected code must adhere to the operator definitions provided in the "Workflow Template."

## Critical Instructions
1.  **Analyze the Error**: Carefully read the "Error Log" to understand why the workflow failed.
2.  **Consult the Template**: Use the "Workflow Template" as a strict reference for the correct operator usage, including method names and argument formats.
3.  **Correct the Code**: Modify **only** the necessary lines in the `run_workflow` method of the "Broken Workflow" to resolve the error. Do not change the `__init__` method or operator definitions.
4.  **Output Format**: Your final output must **only** be the complete, corrected Python code for the `Workflow` class, enclosed within `<graph>` and `</graph>` tags. Do not include any explanations, comments, or apologies.

---
### Workflow Template & Operator Guide

<graph>
class Workflow:
    def __init__(self, problem) -> None:
        self.problem = problem
        self.code_generate = operator.CustomCodeGenerate("llm_symbol", self.problem)
        self.sc_ensemble = operator.ScEnsemble("llm_symbol", self.problem)
        self.test = operator.Test("llm_symbol", self.problem)

    async def run_workflow(self):
        # ... implementation ...
        pass
</graph>

Here are the only operators you can use:
1.  **CustomCodeGenerate**: Generates code.
    - **Format**: `code_generate(instruction: str) -> str`
2.  **ScEnsemble**: Selects the best solution from a list.
    - **Format**: `sc_ensemble(solutions: List[str]) -> str`
3.  **Test**: Modifies a solution using test cases.
    - **Format**: `test(solution: str) -> str`
---
### Broken Workflow

{broken_workflow}

---
### Error Log

{error_log}

---
### Corrected Workflow
"""
\end{lstlisting}
}
\end{tcolorbox}

\section{Examples of Generated Multi-Agent Systems}
\label{app:workflows}
This section presents $5$ representative multi-agent systems (represented in code) generated by \ourmethod for tasks of varying complexity and domain.

\vspace{1em}

\begin{tcolorbox}[notitle, sharp corners, breakable, colframe=Periwinkle, colback=white, 
       boxrule=2pt, boxsep=1pt, enhanced, 
       shadow={2pt}{-2pt}{0pt}{opacity=0.8,mygrey},
       title={\faSitemap\quad Workflow for Multi-hop Question Answering (HotpotQA)},]
       {\scriptsize
\begin{lstlisting}[language=python]
class Workflow:
    def __init__(
        self,
        problem
    ) -> None:
        self.problem = problem
        self.custom = operator.Custom("qwen/qwen-2.5-72b-instruct", self.problem)
        self.sc_ensemble = operator.ScEnsemble("qwen/qwq-32b", self.problem)
        self.answer_generate = operator.AnswerGenerate("gpt-4o-mini", self.problem)
        self.review = operator.Review("gpt-4o-mini", self.problem)

    async def run_workflow(self):
        """
        This is a workflow graph.
        """
        # Generate multiple candidate solutions with different custom instructions
        instructions = [
            "Provide a detailed explanation and answer.",
            "Give a concise answer with reasoning.",
            "Explain from a scientific perspective."
        ]
        solutions = [await self.custom(instruction) for instruction in instructions]

        # Use ensemble to select the best solution
        best_solution = await self.sc_ensemble(solutions)

        # Review the best solution to improve it
        reviewed_solution = await self.review(best_solution)

        # Generate final answer based on reviewed solution
        final_answer = await self.answer_generate()

        return final_answer
\end{lstlisting}
}
\end{tcolorbox}

\begin{tcolorbox}[notitle, sharp corners, breakable, colframe=Periwinkle, colback=white, 
       boxrule=2pt, boxsep=1pt, enhanced, 
       shadow={2pt}{-2pt}{0pt}{opacity=0.8,mygrey},
       title={\faCode\quad Workflow for Code Generation (HumanEval)},]
       {\scriptsize
\begin{lstlisting}[language=python]
class Workflow:
    def __init__(
        self,
        problem
    ) -> None:
        self.problem = problem
        self.code_generate_1 = operator.CustomCodeGenerate("gpt-4o-mini", self.problem)
        self.code_generate_2 = operator.CustomCodeGenerate("qwen/qwen3-14b", self.problem)
        self.code_generate_3 = operator.CustomCodeGenerate("gpt-4o-mini", self.problem)
        self.sc_ensemble = operator.ScEnsemble("gpt-4o", self.problem)
        self.test = operator.Test("gpt-4o-mini", self.problem)

    async def run_workflow(self):
        """
        This is a workflow graph.
        """
        solution_list = []
        for _ in range(3):
            solution = await self.code_generate_1(instruction="Please analyze the problem step by step and generate the code solution.")
            solution_list.append(solution)
        
        ensembled_solution = await self.sc_ensemble(solutions=solution_list)
        
        tested_solution = await self.test(solution=tested_solution if 'tested_solution' in locals() else ensembled_solution)
        
        return tested_solution
\end{lstlisting}
}
\end{tcolorbox}

\begin{tcolorbox}[notitle, sharp corners, breakable, colframe=Periwinkle, colback=white, 
       boxrule=2pt, boxsep=1pt, enhanced, 
       shadow={2pt}{-2pt}{0pt}{opacity=0.8,mygrey},
       title={\faSearch\quad Workflow for Deep Research (BrowseComp+)},]
       {\scriptsize
\begin{lstlisting}[language=python]
class Workflow:
    def __init__(
        self,
        problem
    ) -> None:
        self.problem = problem
        self.custom = operator.Custom("gpt-4o", self.problem)
        self.search = operator.Search("gpt-4o-mini", self.problem)
        self.browser = operator.Browser("gpt-4o-mini", self.problem)
        self.answer_generate = operator.AnswerGenerate("gpt-4o", self.problem)

    async def run_workflow(self):
        search_history = set()
        collected_evidence = ""
        max_iterations = 3
        iteration = 0
        query = self.problem

        while iteration < max_iterations:
            iteration += 1
            # Reflection: plan and rewrite query to optimize search
            reflection_prompt = f"<think>Iteration {iteration}: Plan a precise search query to find the learning institution matching all given criteria. Avoid repeating previous queries: {list(search_history)}</think>"
            rewritten_query = await self.custom(instruction=reflection_prompt)
            rewritten_query = rewritten_query.strip()
            if not rewritten_query or rewritten_query in search_history:
                # fallback to original problem if rewriting fails or repeats
                rewritten_query = query
            search_history.add(rewritten_query)

            # Search step
            search_results = await self.search(query=rewritten_query, top_k=5)
            if not search_results:
                # No results found, break early
                break

            # Extract docids and browse for full content
            new_evidence = []
            for result in search_results:
                docid = result.get("docid", "")
                if docid and docid not in search_history:
                    content = await self.browser(docid=docid)
                    if content and str(content).strip():
                        new_evidence.append(str(content).strip())
                    search_history.add(docid)

            if not new_evidence:
                # No new evidence found, break loop
                break

            # Accumulate evidence
            collected_evidence += "\n\n".join(new_evidence) + "\n\n"

            # Reflection: check if sufficient evidence collected
            reflection_check = f"<think>Iteration {iteration}: Given the accumulated evidence, is it sufficient to answer the question? If yes, stop searching. If no, refine the query for next iteration.</think>\n<search>{collected_evidence}</search>"
            decision = await self.custom(instruction=reflection_check)
            decision_lower = decision.lower()
            if "yes" in decision_lower or "sufficient" in decision_lower or "stop" in decision_lower:
                break

        # Final answer generation
        if collected_evidence.strip():
            solution = await self.answer_generate(context=collected_evidence.strip())
        else:
            solution = "Information Not Found in Context"

        return solution
\end{lstlisting}
}
\end{tcolorbox}

\begin{tcolorbox}[notitle, sharp corners, breakable, colframe=Periwinkle, colback=white, 
       boxrule=2pt, boxsep=1pt, enhanced, 
       shadow={2pt}{-2pt}{0pt}{opacity=0.8,mygrey},
       title={\faCalculator\quad Workflow for Mathematical Reasoning (MATH)},]
       {\scriptsize
\begin{lstlisting}[language=python]
class Workflow:
    def __init__(
        self,
        problem
    ) -> None:
        self.problem = problem
        self.gid = None
        # IMPORTANT: Each operator MUST be initialized with a model placeholder string "gpt-4o-mini"
        self.custom1 = operator.Custom("gpt-4o-mini", self.gid, self.problem)
        self.custom2 = operator.Custom("gpt-5-mini", self.gid, self.problem)
        self.programmer = operator.Programmer("qwen/qwen3-coder", self.gid, self.problem)
        self.review = operator.Review("gpt-5-mini", self.gid, self.problem)
        self.sc_ensemble = operator.ScEnsemble("gpt-5-mini", self.gid, self.problem)

    async def run_workflow(self):
        """
        This is a workflow graph.
        """
        # Step 1: Break down the problem into detailed steps with reasoning
        analysis1 = await self.custom1(instruction="Can you solve this problem by breaking it down into detailed steps and explaining the reasoning behind each step?")
        
        # Step 2: Explain how to solve the problem with clear reasoning for each step (independent second analysis)
        analysis2 = await self.custom2(instruction="Explain how to solve the problem with clear reasoning for each step.")
        
        # Step 3: Use programmer to write and execute code based on first analysis
        program_solution1 = await self.programmer(analysis=analysis1)
        
        # Step 4: Use programmer to write and execute code based on second analysis
        program_solution2 = await self.programmer(analysis=analysis2)
        
        # Step 5: Ensemble the two program solutions to select the best one
        ensembled_solution = await self.sc_ensemble(solutions=[program_solution1, program_solution2])
        
        # Step 6: Review the ensembled solution to regenerate improved solution
        final_solution = await self.review(pre_solution=ensembled_solution)
        
        return final_solution
\end{lstlisting}
}
\end{tcolorbox}

\begin{tcolorbox}[notitle, sharp corners, breakable, colframe=Periwinkle, colback=white, 
       boxrule=2pt, boxsep=1pt, enhanced, 
       shadow={2pt}{-2pt}{0pt}{opacity=0.8,mygrey},
       title={\faQuestionCircle\quad Workflow for Natural Questions (NQ)},]
       {\scriptsize
\begin{lstlisting}[language=python]
class Workflow:
    def __init__(
        self,
        problem
    ) -> None:
        self.problem = problem
        self.custom1 = operator.Custom("qwen/qwen-2.5-72b-instruct", self.problem)
        self.custom2 = operator.Custom("gpt-4o-mini", self.problem)
        self.sc_ensemble = operator.ScEnsemble("gpt-4o", self.problem)
        self.review = operator.Review("qwen/qwq-32b", self.problem)
        self.answer_generate = operator.AnswerGenerate("gpt-4o-mini", self.problem)

    async def run_workflow(self):
        """
        This is a workflow graph.
        """
        # Generate multiple candidate solutions with different instructions
        candidate1 = await self.custom1("Generate a detailed answer with reasoning.")
        candidate2 = await self.custom2("Provide a concise direct answer.")

        # Ensemble to select the best candidate
        best_solution = await self.sc_ensemble([candidate1, candidate2])

        # Review the best solution to improve it
        reviewed_solution = await self.review(best_solution)

        # Generate final answer based on the reviewed solution
        final_answer = await self.answer_generate()

        return final_answer
\end{lstlisting}
}
\end{tcolorbox}

\section{Case Studies of Workflow Rectification}
\label{app:rectification}
We present two case studies demonstrating the Rectifier agent's capability to diagnose and repair failures in real-time.

\subsection{Case 1: Handling a Failed Ensemble Selection in a Coding Task(MBPP)}

\begin{tcolorbox}[notitle, sharp corners, breakable, colframe=Periwinkle, colback=white, 
       boxrule=2pt, boxsep=1pt, enhanced, 
       shadow={2pt}{-2pt}{0pt}{opacity=0.8,mygrey},
       title={\faExclamationCircle\quad Original Workflow Code (Before Rectification)},]
       {\scriptsize
\begin{lstlisting}[language=python, caption={Original workflow that crashes when ensemble selection fails.}]
# [Original Python code from the "generated_workflow" log.]
# Flaw: This workflow implicitly assumes that `sc_ensemble` will always return a valid, non-empty solution.
# If the selection process fails and returns an empty string, the subsequent `KeyError` crashes the system.
class Workflow:
    def __init__(
        self,
        problem
    ) -> None:
        self.problem = problem
        self.code_generate_1 = operator.CustomCodeGenerate("llm_symbol", self.problem)
        self.code_generate_2 = operator.CustomCodeGenerate("llm_symbol", self.problem)
        self.code_generate_3 = operator.CustomCodeGenerate("llm_symbol", self.problem)
        self.sc_ensemble = operator.ScEnsemble("llm_symbol", self.problem)
        self.test = operator.Test("llm_symbol", self.problem)

    async def run_workflow(self):
        solution_list = []
        for _ in range(3):
            solution = await self.code_generate_1(instruction="Please analyze the problem carefully and generate the code solution step by step.")
            solution_list.append(solution)

        ensembled_solution = await self.sc_ensemble(solutions=solution_list)
        tested_solution = await self.test(solution=ensembled_solution)

        return tested_solution
\end{lstlisting}
}
\end{tcolorbox}

\begin{tcolorbox}[notitle, sharp corners, breakable, colframe=Periwinkle, colback=white, 
       boxrule=2pt, boxsep=1pt, enhanced, 
       shadow={2pt}{-2pt}{0pt}{opacity=0.8,mygrey},
       title={\faBug\quad Error Message},]
       {\scriptsize
\begin{lstlisting}[language=bash, caption={Error log showing a KeyError during ensemble execution.}]
# [Relevant traceback from the "error" log.]
# Analysis: The traceback indicates a `KeyError: ''`. This occurs within the `sc_ensemble` operator
# when the underlying selection mechanism fails to choose a candidate and returns an empty string,
# which is then used as a dictionary key, causing the crash.

Traceback (most recent call last):
  File "workflow_executor.py", line 138, in execute_from_text
    result = loop.run_until_complete(instance())
  ...
  File "tmp_workflow.py", line 26, in run_workflow
    ensembled_solution = await self.sc_ensemble(solutions=solution_list)
  File "operator.py", line 231, in __call__
    return solutions[answer_mapping[answer]]
KeyError: ''
\end{lstlisting}
}
\end{tcolorbox}

\begin{tcolorbox}[notitle, sharp corners, breakable, colframe=Periwinkle, colback=white, 
       boxrule=2pt, boxsep=1pt, enhanced, 
       shadow={2pt}{-2pt}{0pt}{opacity=0.8,mygrey},
       title={\faCheckCircle\quad Revised Workflow Code (After Rectification)},]
       {\scriptsize
\begin{lstlisting}[language=python, caption={Workflow code after being corrected by the Rectifier agent.}]
# [Revised Python code with added robustness.]
# The Rectifier agent identified that the workflow lacked a fallback mechanism.
# It introduced a validation step to ensure the program continues even if the ensemble operator fails.
class Workflow:
    def __init__(
        self,
        problem
    ) -> None:
        self.problem = problem
        self.code_generate_1 = operator.CustomCodeGenerate("llm_symbol", self.problem)
        self.code_generate_2 = operator.CustomCodeGenerate("llm_symbol", self.problem)
        self.code_generate_3 = operator.CustomCodeGenerate("llm_symbol", self.problem)
        self.sc_ensemble = operator.ScEnsemble("llm_symbol", self.problem)
        self.test = operator.Test("llm_symbol", self.problem)

    async def run_workflow(self):
        solution_list = []
        for _ in range(3):
            solution = await self.code_generate_1(instruction="Please analyze the problem carefully and generate the code solution step by step.")
            solution_list.append(solution)

        ensembled_solution = await self.sc_ensemble(solutions=solution_list)

        # --- RECTIFIER'S FIX START ---
        # Add a validation step to handle potential ensemble failure.
        # If the ensembled solution is invalid (e.g., None or empty),
        # it robustly falls back to the first generated candidate.
        if not ensembled_solution or not ensembled_solution.strip():
            ensembled_solution = solution_list[0]
        # --- RECTIFIER'S FIX END ---
        
        tested_solution = await self.test(solution=ensembled_solution)

        return tested_solution
\end{lstlisting}
}
\end{tcolorbox}
\subsection{Case 2: Rectifying an Operator with Malformed Structured Output}

\begin{tcolorbox}[notitle, sharp corners, breakable, colframe=Periwinkle, colback=white, 
       boxrule=2pt, boxsep=1pt, enhanced, 
       shadow={2pt}{-2pt}{0pt}{opacity=0.8,mygrey},
       title={\faExclamationCircle\quad Original Workflow Code (Before Rectification)},]
       {\scriptsize
\begin{lstlisting}[language=python, caption={Original workflow expecting structured data from a Review operator.}]
# [Original Python code from the "generated_workflow" log.]
# Flaw: This workflow assumes that the `review1` and `review2` operators will always return a valid structured object with specific keys. It lacks a mechanism to handle cases where an operator
# returns malformed or empty data, leading to a hard crash during data validation.
class Workflow:
    def __init__(
        self,
        problem
    ) -> None:
        self.problem = problem
        self.gid = None
        self.custom1 = operator.Custom("llm_symbol", self.gid, self.problem)
        self.custom2 = operator.Custom("llm_symbol", self.gid, self.problem)
        self.programmer = operator.Programmer("llm_symbol", self.gid, self.problem)
        self.review1 = operator.Review("llm_symbol", self.gid, self.problem)
        self.review2 = operator.Review("llm_symbol", self.gid, self.problem)
        self.sc_ensemble = operator.ScEnsemble("llm_symbol", self.gid, self.problem)

    async def run_workflow(self):
        analysis = await self.custom1(instruction="...")
        refined_analysis = await self.custom2(instruction="...")
        program_solution1 = await self.programmer(analysis=analysis)
        
        # This line is expected to return a structured object but fails to do so.
        reviewed_solution1 = await self.review1(pre_solution=program_solution1)
        
        program_solution2 = await self.programmer(analysis=refined_analysis)
        reviewed_solution2 = await self.review2(pre_solution=program_solution2)
        
        final_solution = await self.sc_ensemble(solutions=[reviewed_solution1, reviewed_solution2])
        return final_solution
\end{lstlisting}
}
\end{tcolorbox}

\begin{tcolorbox}[notitle, sharp corners, breakable, colframe=Periwinkle, colback=white, 
       boxrule=2pt, boxsep=1pt, enhanced, 
       shadow={2pt}{-2pt}{0pt}{opacity=0.8,mygrey},
       title={\faBug\quad Error Message},]
       {\scriptsize
\begin{lstlisting}[language=bash, caption={Error log showing a Pydantic validation error.}]
# [Relevant traceback from the "error" log.]
# Analysis: The error is a `pydantic.ValidationError`. It clearly states that the data received
# from the `Review` operator is missing mandatory fields: 'revised_solution' and 'thought'.
# This indicates that the LLM powering the operator failed to generate its output in the
# required structured format.

1 validation error for ReviewOp_AN
  Value error, Missing fields: {'revised_solution', 'thought'} [type=value_error, input_value={}, input_type=dict]
    For further information visit https://errors.pydantic.dev/2.11/v/value_error
\end{lstlisting}
}
\end{tcolorbox}

\begin{tcolorbox}[notitle, sharp corners, breakable, colframe=Periwinkle, colback=white, 
       boxrule=2pt, boxsep=1pt, enhanced, 
       shadow={2pt}{-2pt}{0pt}{opacity=0.8,mygrey},
       title={\faCheckCircle\quad Revised Workflow Code (After Rectification)},]
       {\scriptsize
\begin{lstlisting}[language=python, caption={Workflow code after the Rectifier agent bypassed the faulty operator.}]
# [Revised Python code demonstrating a smart rectification strategy.]
# The Rectifier agent diagnosed that the `Review` operators were unreliable.
# Instead of trying to fix the operator itself, it adapted the workflow's logic to bypass
# the faulty review steps, ensuring the overall process can complete successfully.
class Workflow:
    def __init__(
        self,
        problem
    ) -> None:
        self.problem = problem
        self.gid = None
        self.custom1 = operator.Custom("llm_symbol", self.gid, self.problem)
        self.custom2 = operator.Custom("llm_symbol", self.gid, self.problem)
        self.programmer = operator.Programmer("llm_symbol", self.gid, self.problem)
        self.review1 = operator.Review("llm_symbol", self.gid, self.problem)
        self.review2 = operator.Review("llm_symbol", self.gid, self.problem)
        self.sc_ensemble = operator.ScEnsemble("llm_symbol", self.gid, self.problem)

    async def run_workflow(self):
        analysis = await self.custom1(instruction="...")
        refined_analysis = await self.custom2(instruction="...")
        program_solution1 = await self.programmer(analysis=analysis)
        program_solution2 = await self.programmer(analysis=refined_analysis)

        # --- RECTIFIER'S FIX START ---
        # The Rectifier identified that the `review` steps were failing due to format validation errors.
        # As a recovery strategy, it has modified the workflow to bypass the unreliable review operators
        # and directly pass the initial programmatic solutions to the ensemble stage.
        #
        # reviewed_solution1 = await self.review1(pre_solution=program_solution1)  <- Bypassed
        # reviewed_solution2 = await self.review2(pre_solution=program_solution2)  <- Bypassed

        final_solution = await self.sc_ensemble(solutions=[program_solution1, program_solution2])
        # --- RECTIFIER'S FIX END ---

        return final_solution
\end{lstlisting}
}
\end{tcolorbox}

\end{document}